\documentclass[prl,aps,twocolumn,showpacs]{revtex4}
\usepackage{graphicx,amsmath,amsfonts,latexsym,bm,eso-pic}
\usepackage{color}
\usepackage{soul}
\usepackage{verbatim}

\def \beq{\begin{equation}}
\def \eeq{\end{equation}}
\def \beqa{\begin{eqnarray}}
\def \eeqa{\end{eqnarray}}
\def \sx{\sigma_x}

\def \sz{\sigma_z}
\def \tx{\tau_x}

\def \tz{\tau_z}

\newcommand{\rem}[1]{}
\newcommand{\refe}[1]{(\ref{#1})}
\newcommand{\fige}[1]{Fig.~\ref{#1}}
\newcommand{\refE}[1]{Eq.~(\ref{#1})}

\def\eq{&=&} 

\begin{document}

\title{Dynamics of Majorana States in a Topological Josephson Junction}

\author{Manuel Houzet}
\author{Julia S. Meyer}
\author{Driss M. Badiane}
\affiliation{SPSMS, UMR-E 9001 CEA/UJF-Grenoble 1, INAC, Grenoble, F-38054, France}
\author{Leonid I. Glazman} 
\affiliation{Department of Physics and Applied Physics, Yale University, New Haven, Connecticut 06520, USA}
\date{\today}



\begin{abstract}
Topological Josephson junctions carry $4\pi$-periodic bound states. A finite bias applied to the junction limits the lifetime of the bound state by dynamically coupling it to the continuum. Another characteristic time scale, the phase adjustment time, is determined by the  resistance of the circuit \lq\lq seen\rq\rq\ by the junction. We show that the $4\pi$ periodicity manifests itself by an even-odd effect in Shapiro steps only if the phase adjustment time is shorter than the lifetime of the bound state. The presence of a peak in the current noise spectrum at half the Josephson frequency is a more robust manifestation of the $4\pi$ periodicity, as it persists for an arbitrarily long phase adjustment time. We specify, in terms of the circuit parameters, the conditions necessary for observing the manifestations of $4\pi$ periodicity in the noise spectrum and Shapiro step measurements.
\end{abstract}

\maketitle

Topological Josephson junctions have attracted much interest lately as a means of probing the zero-energy Majorana fermion states that exist at the surface of topological superconductors. Such topological superconductors may be realized via the proximity effect by combining conventional superconductors with two-dimensional (2D) topological insulators \cite{fu2009} or with nanowires in the presence of both strong spin-orbit coupling and a magnetic field \cite{majorana1,majorana2}. Recently several experiments have reported evidence of zero-energy states in nanowire-based systems 
\cite{Kouwenhoven,Heiblum,Rokhinson}.
To confirm their Majorana nature, additional experimental signatures are desirable.

In a topological Josephson junction, the Majorana bound states localized on either side of the junction hybridize and form an Andreev bound state whose energy $\epsilon_A(\varphi)$ is $4\pi$ periodic in the phase difference $\varphi$ between the two superconductors. Depending whether the state is occupied or empty, the energy of the junction is $\pm\epsilon_A(\varphi)/2$. In the presence of parity-changing processes, the occupation of the state may change. Thus, the equilibrium Josephson current displays the usual $2\pi$ periodicity as the system follows the ground state. By contrast, if upon phase variation the system follows one branch of the spectrum, then $4\pi$ periodicity should appear indeed. As a result, under dc bias voltage $V_{\rm dc}$, such a system has been predicted to manifest a fractional ac Josephson effect  \cite{kitaev2001,kwon2003,fu2009} at frequency $\omega_J/2=eV_{\rm dc}/\hbar$, that is at half of the \lq\lq usual\rq\rq\ Josephson frequency. By the same token, in the presence of an additional ac bias with frequency $\Omega$, one would expect an even-odd effect: namely only the even Shapiro steps at $eV_{\rm dc}=k\Omega$ ($k\in\mathbb{Z}$) should be visible in the current-voltage characteristics~\cite{kwon2003,eoShapiro,Dominguez}. However, the application of a bias voltage inevitably couples the bound state to the continuum, thus causing its occupation to switch. The corresponding switching rate determines the lifetime of the bound state, $\tau_s$.
In addition to these intrinsic processes, the evolution of the phase difference across the junction depends on the properties of the circuit connecting the Josephson junction to the voltage source. Any nonzero resistance ${\cal R}$ of the connection allows for an adjustment of the phase difference  over some characteristic time~\cite{Shapiro,kautz} $\tau_R\propto {\cal R}^{-1}$.

In this work, we evaluate the lifetime of Majorana  bound states, $\tau_s$, limited by their dynamic coupling to the continuum. This mechanism gains importance in nearly ballistic junctions and leads to a strong dependence of $\tau_s$ on the applied voltage. We show that the transport properties of the junction crucially depend on two characteristic time scales, $\tau_s$ and $\tau_R$. If $\tau_s\gg\tau_R$,  Majorana states lead to an even-odd effect in the height of Shapiro steps, in agreement with Refs. \cite{kwon2003,eoShapiro,Dominguez}. By contrast, if $\tau_s\ll\tau_R$, all Shapiro steps are suppressed. However, signatures of the $4\pi$ periodicity are still visible in the noise spectrum which, under dc {voltage bias},  displays peaks at $\omega=\pm\omega_J/2$, as was seen in  numerical simulations~\cite{Badiane2011}. Here we develop an analytical theory for the noise spectrum and find the dependence of the peak widths on $\tau_s$. While noise measurements in the gigahertz range are not easy to realize, the low-frequency noise is more accessible. The down-conversion of the noise peak to $\omega=\pm(\omega_J/2-\Omega)$ may be achieved by adding a small ac bias of frequency $\Omega$.

{To examine the nonadiabatic transitions between the Majorana state and quasiparticles continuum, we} consider the helical edge state of a 2D topological insulator in which superconductivity has been induced by two superconducting contacts in order to create a topological Josephson junction of length $L$.
The system is described by the Hamiltonian
\beq
{\cal H}=v p \sz \tz-eU(x,t)\tz+M(x) \sx+\Delta(x)e^{i\phi(x,t)\tz}\tx.
\label{eq:H}
\eeq
Here, $v$ is the Fermi velocity, $p$ is the momentum operator, $U(x,t)=V(t)[\theta(-x)-\theta(x-L)]/2$ is the electric potential, $M(x)=M\theta(x)\theta(L-x)$ is a transverse magnetic field within the junction, $\Delta(x)=\Delta[\theta(-x)+\theta(x-L)]$ and $\phi(x,t)=\varphi(t)[\theta(-x)-\theta(x-L)]/2$, with $\dot\varphi(t)=2eV(t)$, are the amplitude and phase of the superconducting order parameter in the left and right leads, and $\sigma_i$, $\tau_j$ ($i,j=x,y,z$) are Pauli matrices acting in the spin and particle-hole spaces, respectively. All energies are measured from the chemical potential. 

We concentrate on the case of a short junction, $L\ll\xi$, where $\xi=v/\Delta$ is the superconducting coherence length. In equilibrium ($V=0$), such a junction hosts a single Andreev bound state with energy 
\beq
\epsilon_A(\varphi)=\sqrt D \Delta \cos(\varphi/2),
\label{eq:EA}
\eeq
where $D$ is the transmission probability of the junction which depends on its length and on the magnitude of the transverse field. Thus, the minimal gap $\delta$ between the bound state and the continuum at $\varphi= 2n\pi\enspace(n\in\mathbb{Z})$ is given as $\delta=\Delta(1-\sqrt D)$. In the following we consider a highly transmitting junction where $\delta\approx \Delta R/2$ and $R=1-D\sim (ML/v)^2$ is the reflection probability. 

Out of equilibrium, nonadiabatic transitions between the Andreev bound state and the continuum are induced. These transitions change the occupation of the bound state and thus lead to switching between the two current branches, $I(\varphi)=\pm I_J\sin(\varphi/2)$, where $I_J=e\sqrt{D}\Delta/2$. At dc bias $eV_{\rm dc}\ll\Delta$, violation of the adiabaticity occurs in narrow intervals $|\varphi-2\pi n|\ll \pi$ of the time-varying phase $\varphi=2eV_{\rm dc}t$. 
To find the corresponding probability of a nonadiabatic transition between the localized Majorana state and continuum, we concentrate on 
the case $n=0$, corresponding to the time interval $|t|\ll\pi/(eV_{\rm dc})$. Using a gauge transformation ${\cal H}\to \tilde{\cal H}=U^\dagger{\cal H}U-iU^\dagger \dot U$ with $U=\exp[i\phi\tz/2]$ and taking the limit $L\to 0$ (keeping $R$ fixed), we obtain for the said interval of $\varphi$ the simplified Hamiltonian $\tilde{\cal H}= vp\sz\tz+\Delta\tx +v[(\varphi/2)\sz+\sqrt R\sx]\delta(x)$. Furthermore, at $eV_{\rm dc}\ll\Delta$, only states close to the continuum edge, $v|p|\ll\Delta$, are relevant. Thus, after diagonalizing the bulk Hamiltonian, we can restrict ourselves to  a $2\times2$ subspace of the initial spin and particle-hole space,
\beq
H=\Delta +\frac{v^2p^2}{2\Delta} +v\left(\frac12\varphi \sz+\sqrt R \sx\right)\delta(x).
\label{eq:Hn}
\eeq
Equation \eqref{eq:Hn} describes a spin-degenerate continuum with quadratic dispersion, in the presence of a spin-dependent local potential.  
The first term in this potential accounts for the phase shift across the barrier in a gauge with zero electric potential in the leads and a vector potential localized at the barrier. The second term describes the magnetic barrier. 

For a fixed phase, the Hamiltonian \refe{eq:Hn} accommodates a single bound state with energy $\epsilon_A(\varphi)=\Delta\left(1-\varphi^2/8-R/2\right)$, in agreement with \refE{eq:EA} at $R,\varphi^2\ll 1$. A particle occupying this bound state at time $t\rightarrow-\infty$ (within the simplified model) has a probability $s$ to 
escape to the continuum as the phase increases.
The problem is, thus, a generalization to a two-band model of the transition from a discrete state to a continuum, considered by Demkov and Osherov \cite{DemkovOsherov1967}. Dimensional analysis shows that the transition (or switching) probability
is determined by the adiabaticity parameter $\lambda=R^{3/2}\Delta/(eV_{\rm dc})$. Below we find this probability in two limiting cases of the parameter $\lambda$.

Let us start with the antiadiabatic regime, $\lambda\ll 1$. At $\lambda=0$ the spin bands in \refE{eq:Hn} are decoupled. At times $t<0$, the bound state belongs to the spin-up band whereas, at times $t>0$, the bound state belongs to the spin-down band. The spin-up bound state is described by a wave function $|\psi_\uparrow(t)\rangle$ whose projection on the position of the local potential is
\beq
\langle x=0|\psi_\uparrow(t)\rangle=\frac {\tau} {\sqrt{2\pi \ell }}\int_{\cal C}d\omega\; e^{i\omega t}\;e^{i(-2\omega\tau)^{3/2}/3}|\uparrow\rangle.
\label{eq:wavefuntion}
\eeq
The wave function $|\psi_\downarrow(t)\rangle$ of the spin-down bound state is related to $|\psi_\uparrow(t)\rangle$ by time reversal. Here, the characteristic length and time scales are given by $\ell=v/[\Delta^2eV_{\rm dc}]^{1/3}$ and $\tau=1/[\Delta(eV_{\rm dc})^2]^{1/3}$, respectively. Furthermore, ${\cal C}$ is a contour in the complex $\omega$ plane \cite{DemkovOsherov1967} that starts and ends at infinity, with arguments $\pi<\theta<5\pi/3$ and $0<\theta<\pi/3$, respectively, and avoids the branch cut along the positive real axis. As the spin bands are decoupled, a particle occupying the (spin-up) bound state at  $t=-\infty$, has probability $1-s=0$ to occupy the (spin-down) bound state as $t\to\infty$.

A finite $\lambda$ couples the two bands and, thus, enables spin flips. The switching probability
$s$ can be obtained from the overlap $c_\downarrow(t)=\langle\psi_\downarrow(t)|\psi(t)\rangle$ of the exact wave function, $|\psi(t)\rangle$, where $|\psi(-\infty)\rangle=|\psi_\uparrow(-\infty)\rangle$, with the wave function of the spin-down bound state, $|\psi_\downarrow(t)\rangle$, through  $s =1-|c_\downarrow(\infty)|^2$. At $\lambda\ll 1$, $c_\downarrow$ can be computed perturbatively using
\beq
\dot c_\downarrow(t) 
\approx 
i
\langle \psi_\downarrow(t)|v\sqrt{R}\sx\delta(x)|\psi_\uparrow(t)\rangle.
\label{eq:eqdiff}
\eeq
Solving the differential equation \refe{eq:eqdiff} to obtain $c_\downarrow(\infty)$ and computing $s$, we find  $s\approx 1-1.05\,\lambda^{2/3}$. 
The time scale over which the transition happens is $\tau_\mathrm{t}\sim\tau$.

In the quasiadiabatic regime, $\lambda\gg 1$, it is convenient to expand the exact wave function 
in the adiabatic basis of \refE{eq:Hn}, 
\beq
|\psi(t)\rangle=c_A(t)|\psi_A(t)\rangle+\sum_{p\sigma}c_{p\sigma}(t)|\psi_{p\sigma}(t)\rangle.
\label{eq:expansion}
\eeq
Here $|\psi_A(t)\rangle$ and $|\psi_{p\sigma}(t)\rangle$ are the adiabatic wave functions for the bound state and the doubly degenerate states of the continuum, respectively. 
[Note that $|\psi_A(\mp\infty)\rangle=|\psi_{\uparrow,\downarrow}(\mp\infty)\rangle$.]
The switching probability
$s$ is related to the amplitudes $c_{p\sigma}(\infty)$ of the continuum states in \refE{eq:expansion} through $s=\sum_{p\sigma}\left|c_{p\sigma}(\infty)\right|^2$, using the initial conditions $c_A(-\infty)=1$ and $c_{p\sigma}(-\infty)=0$.
At $\lambda\gg1$, using
\beq
\dot c_{p\sigma}(t)
\approx 
i
\dot\varphi(t)\,
\frac{\langle\psi_{p\sigma}(t)|\frac{\partial H}{\partial\varphi(t)}|\psi_A(t)\rangle}{\epsilon_{p\sigma}-\epsilon_A(t)}e^{-i\int^tds\,[\epsilon_{p\sigma}-\epsilon_A(s)]}\ ,
\eeq
the amplitudes $c_{p\sigma}(\infty)$ are expressed through integrals that may be evaluated by a saddle point method. We obtain 
$s\simeq 0.93\,\lambda^{-5/4}e^{-2\lambda/3}$.
Furthermore, we can identify the time scale over which the transition happens, $\tau_\mathrm{t}\sim\sqrt R/(eV_{\rm{dc}})$.

\begin{figure}
\includegraphics[width=0.8\linewidth]{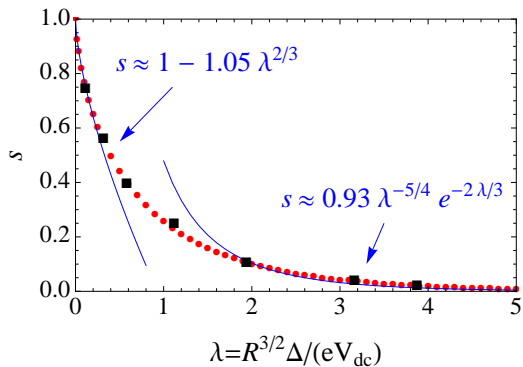}
\caption{
Switching probability $s$ as a function of the adiabaticity parameter $\lambda$. Dots: $s$ found from a numerical solution of the Schr\"odinger equation with Hamiltonian \refe{eq:Hn}. Lines: asymptotic expressions for $s$, see text. Squares: $s$ extracted from the \lq\lq brute-force\rq\rq 
evaluation of the noise spectrum by solving numerically the problem of multiple Andreev reflections \cite{Badiane2011} and fitting the result by Eq. \refe{eq:Svsomega}; see \cite{SM} for details.
}
\label{fig-switch}
\end{figure}

At arbitrary $\lambda$, the switching probability
can be obtained numerically by discretizing \refE{eq:Hn} on a tight-binding lattice and solving the corresponding Schr\"{o}dinger equation numerically. The result, together with the asymptotes obtained above, is shown in  \fige{fig-switch}.

\begin{figure}
(a)\quad\includegraphics[width=0.75\linewidth]{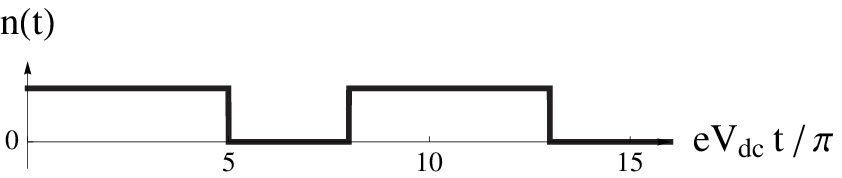}\\[0.25cm]
(b)\quad\includegraphics[width=0.75\linewidth]{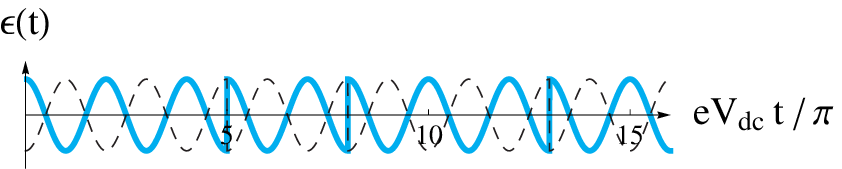}\\[0.25cm]
(c)\quad\includegraphics[width=0.75\linewidth]{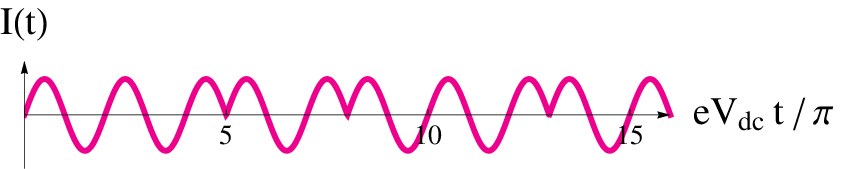}
\caption{Schematic view of the switching processes due to the coupling with the continuum: (a) occupation $n$ of the bound state, (b) energy $\epsilon$ of the system, and (c) Josephson current $I$ as a function of time $t$ under dc bias voltage $V_{\rm{dc}}$.}
\label{fig-ABS}
\end{figure}

Using the fact that the transition time $\tau_\mathrm{t}$ is much shorter than the 
Josephson oscillation period, $\tau_\mathrm{t}\ll\pi/(eV_\mathrm{dc})$,
we may now write an effective discrete Markov model for the bound state dynamics; cf. Fig. \ref{fig-ABS}. 
Using a discrete time evolution we assume that if the state is filled, at phase $\varphi_{2n}=4n\pi$, there is a probability $s$ of the particle to 
escape from the bound state to the continuum,
whereas if the state is empty, at time $\varphi_{2n+1}=(4n+2)\pi$, there is a probability $s$ of a particle from the continuum filling the bound state. Thus,
\begin{subequations}
\begin{eqnarray}
\begin{pmatrix}P_{2n+1}\\Q_{2n+1}\end{pmatrix}\eq\begin{pmatrix}1&s\\0&1-s\end{pmatrix}\begin{pmatrix}P_{2n}\\Q_{2n}\end{pmatrix},\\
\begin{pmatrix}P_{2n}\\Q_{2n}\end{pmatrix}\eq\begin{pmatrix}1-s&0\\s&1\end{pmatrix}\begin{pmatrix}P_{2n-1}\\Q_{2n-1}\end{pmatrix}.
\end{eqnarray} 
\end{subequations}
Here $P_n$ is the probability for the state to be occupied, and $Q_n=1-P_n$ is the probability for the state to be empty at phases $\varphi_n<\varphi(t)<\varphi_{n+1}$, corresponding to $n={\rm Int}\left[{\varphi(t)}/{(2\pi)}\right]$. Solving these equations iteratively, we obtain
\begin{subequations}
\label{eq-proba}
\begin{eqnarray}
P_{2(n+k)}\eq P^\infty_{2n}+(1\!-\!s)^{2k}\left(P_{2n}-P^\infty_{2n}\right),\\
P_{2(n+k)+1}\eq P^\infty_{2n+1}+(1\!-\!s)^{2k+1}\left(P_{2n}-P^\infty_{2n}\right)\!.\qquad
\end{eqnarray} 
\end{subequations}
The long-time probabilities at $k\gg-1/\ln(1-s)$, corresponding to $t\gg\tau_s=-2\pi/[eV_{\rm dc}\ln(1-s)]$, are $4\pi$ periodic and independent of the initial state:
$P_{n}^\infty=[1-(-1)^n s/(2-s)]/2$.
 
In order to determine the transport properties of the junction, the switching time $\tau_s$ has to be compared to other characteristic time scales of the system. In particular, if the junction is embedded into a circuit with a resistance ${\cal R}$ in series, the phase difference across the junction may adjust over a typical time scale $\tau_R\propto {\cal R}^{-1}$.

If $\tau_s\gg\tau_R$, switching may be neglected (i.e., the current may be obtained using the initial occupations $P_0/Q_0$). Computing the dc current in the presence of an applied voltage $V(t)=V_{\rm dc}+V_{\rm ac}\cos(\Omega t)$ with $V_{\rm ac}\ll V_{\rm dc}$ then yields the even-odd effect discussed in Refs.~\cite{kwon2003,eoShapiro,Dominguez}. Namely, taking into account a finite resistance ${\cal R}$, the average current reads
\begin{equation}
\label{eq-current-RSJ}
I_{\rm dc}=\sum_k \frac{\delta V_k}{{\cal R}}\left\{1-\theta\left[1-\left(\frac{{\cal R}I_k}{\delta V_k}\right)^2\right]\sqrt{1-\left(\frac{{\cal R}I_k}{\delta V_k}\right)^2}\right\},
\end{equation}
where $I_k=I_J|J_{k}(\alpha)|$ is the height of the Shapiro step at $eV_{\rm dc}=k\Omega$ and $\delta V_k=V_{\rm dc}-k\Omega/e$. Here $J_k$ are the Bessel functions and $\alpha=eV_{\rm{ac}}/\Omega$. The characteristic time scale may be identified as~\cite{nonTopo,SM} $\tau_R^{(k)}=1/(e{\cal R}I_k)$ at $eV_{\rm dc}\sim k\Omega$. 

A small resistance satisfying the relation ${\cal R}I_J\ll\Omega$ is advantageous for the resolution of Shapiro steps. Furthermore, long switching times require $s\ll 1$. In the small-$s$ regime, the time scale $\tau_s$ decreases as $\exp[2R^{3/2}\Delta/(3k\Omega)]$ with increasing $k$. On the other hand,
at small ac perturbation, $\alpha\ll 1$, the time scale $\tau_R^{(k)}$ increases exponentially with $k$. The crossover from $\tau_s\gg\tau_R$ to the opposite limit, $\tau_s\ll\tau_R$,  upon increasing $V_{\rm dc}$ may occur without violation of the condition $s\ll 1$. This restricts the number of observable Shapiro steps in the current-voltage characteristics, as we show now. At $\tau_s\ll\tau_R$, we may use the long-time probabilities $P^\infty/Q^\infty$ to compute the current and take the limit $\tau_R\to\infty$. At long times, the average current is $2\pi$ periodic,
\begin{eqnarray}
\langle I(t) \rangle&=&I_J\sin\frac{\varphi(t)}2 \left\{Q^\infty_{{\rm Int}\left[{\varphi(t)}/{(2\pi)}\right]}-P^\infty_{{\rm Int}\left[{\varphi(t)}/{(2\pi)}\right]}\right\}\nonumber\\
&=&\frac{sI_J}{2-s} \left|\sin\frac{\varphi(t)}2\right|.\!\!\!
\label{eq-current2}
\end{eqnarray}
More importantly, Eq.~\eqref{eq-current2} shows that the current is proportional to the switching probability
$s$, when $s\ll1$. 

The result \eqref{eq-current2} remains valid in the presence of microwave irradiation as long as $V_{\rm ac}\ll V_{\rm dc}$ and $\Omega\ll \delta$.
The first condition ensures that the ac bias only weakly perturbs the phase velocity $\dot\varphi$. The second condition ensures that ionization of the Majorana level by the ac perturbation would require the absorption of a large number of photons $\sim \delta/\Omega$ and, thus, has a small probability. Note that, for $s\ll1$, the second condition is always satisfied at $\Omega\sim eV_{\rm{dc}}$. Then, $s$ may be approximated by its value at dc bias only.  As a consequence, Eq.~\eqref{eq-current2} implies that Shapiro steps are strongly suppressed, $\langle I_k \rangle\propto s$.

In order to  reveal signatures of the $4\pi$ periodicity in the regime $\tau_s\ll\tau_R$,
we now turn to the current noise spectrum,
\begin{equation}
S(\omega)=2\int\limits_{0}^\infty \!d\tau\cos(\omega\tau)
\overline{\langle\delta I(t)\delta I(t+\tau)\rangle},
\label{eq-noise}
\end{equation}
where $\delta I=I-\langle I \rangle$ and the bar denotes time averaging.
It may be obtained from the correlator 
$
\langle I(\varphi_1) I(\varphi_2)\rangle
= 
I_J^2\sin({\varphi_1}/2)\sin(\varphi_2/2)
[
Q_{n_1}^\infty x_{n_2}(P_{n_1}\!=\!0)
-P_{n_1}^\infty x_{n_2}(P_{n_1}\!=\!1)]$
at $\varphi_1<\varphi_2$,
where $n_i={\rm Int}[\varphi_i/(2\pi)]$. Using the conditional probabilities obtained from Eqs. \eqref{eq-proba}, we find
\beq
\label{eq:correl}
\langle \delta I(\varphi_1)\delta I(\varphi_2)\rangle
= 
\frac{4I_J^2(1-s)}{(2-s)^2}\sin\frac{\varphi_1}2\sin\frac{\varphi_2}2
(1-s)^{n_2-n_1}.
\eeq
At dc bias only, the noise spectral density evaluates to
\begin{equation}
\label{eq:Svsomega}
S(\omega)=\frac{4sI_J^2}{\pi(2-s)} \frac{(eV_{\rm dc})^3}{[\omega^2\!-\!(eV_{\rm dc})^2]^2}\frac{4\cos^2\!\frac{\pi \omega}{2eV_{\rm dc}}}{4\cos^2\!\frac{\pi\omega}{2eV_{\rm dc}}+\frac{s^2}{1- s}}.
\end{equation}
If $s\ll 1$, it has sharp peaks at $\omega=\pm eV_{\rm{dc}}$, i.e., at half of the ``usual" Josephson frequency:
\begin{eqnarray}
S(\omega)&\simeq& \frac{I_J^2}2\frac{seV_{\rm dc}/\pi}{(\omega\mp eV_{\rm dc})^2+(seV_{\rm dc}/\pi)^2}
\label{eq-fin_simp}
\end{eqnarray}
at $|\omega\mp eV_{\rm{dc}}|\ll eV_{\rm{dc}}$. In particular, the peak width is $2seV_{\rm dc}/\pi$. The position of the peak reveals the $4\pi$ periodicity of the Andreev bound state whereas the inverse width characterizes its lifetime $\tau_s\propto s^{-1}$. The peak in the noise is due to the transient $4\pi$-periodic behavior \cite{Aguado} of the current at times smaller than the lifetime of the bound state.

Under microwave irradiation, the peak may be shifted to smaller frequencies. In particular, in the limit $V_{\rm ac}\ll V_{\rm dc},\Omega/e$, we find
\begin{equation}
S(\omega)\simeq \frac{I_J^2}2 J_k^2(\alpha)\frac{s eV_{\rm dc}/\pi }{[\omega\mp(eV_{\rm dc}\!-\!k\Omega)]^2+(s eV_{\rm dc}/\pi)^2}
\label{eq:Sac}
\end{equation} 
at $|\omega\mp(eV_{\rm{dc}}-k\Omega)|\ll eV_{\rm{dc}}$. As above, the peak width is set by the lifetime of the bound state which, thus, may be probed by noise measurements. Equation \eqref{eq:Sac} holds for frequencies $\omega$ not too close to zero. In the limit $\omega\to0$, additional features related to the Shapiro steps may appear \cite{SM}.

While we considered the helical edge states of 2D topological insulators, the model is also applicable to nanowires  \cite{majorana1,majorana2} with strong spin-orbit coupling and a Zeeman energy much larger than $\Delta$. Note that, in addition to the nonadiabatic processes that we considered, nonadiabatic processes in the vicinity of $\varphi=(2n+1)\pi$ become important if the zero-energy crossing is split due to the presence of additional Majorana modes at the ends of the wire \cite{Nazarov,Aguado,Dominguez,footnote2}. In particular, in order to see signatures associated with the $4\pi$ periodicity, the probability of Landau-Zener tunneling across the gap at $\varphi=(2n+1)\pi$ would have to be large while the switching probability due to the coupling with the continuum, discussed in this work, remains small.

To summarize, we analyzed the electron transport through a topological Josephson junction imbedded in a realistic circuit. The Majorana states associated with the junction may lead to two effects, namely (1) an even-odd effect in the Shapiro steps, and (2) a peak in the current noise spectrum at half of the usual Josephson frequency. We found the conditions for these effects to occur. For that we identified the characteristic relaxation time scales for the junction: the lifetime of the bound state originating in its dynamic coupling to the continuum, and the phase adjustment time caused by the resistive environment provided by the circuit. The even-odd effect in the Shapiro steps requires the phase adjustment time to be shorter than the lifetime. For longer phase adjustment times, the even-odd effect is lost. The charactersitic peak in the noise spectrum is less sensitive to the ratio of the two relaxation times. In the limit of long phase adjustment time, the width of the peak provides a measure for the rate of parity-changing processes. The peak at $\omega=eV_{\rm{dc}}/\hbar$ should be seen easily if the dc voltage satisfies the condition $eV_{\rm{dc} }< R^{3/2}\Delta$, where $R$ is the reflection probability. The peak position can be down-shifted in frequency by applying an additional ac bias to the circuit.

In the final stages of preparing the manuscript, we became aware of Ref. \cite{arxiv} considering related effects in nanowire-based topological Josephson junctions.

\acknowledgements

We would like to acknowledge helpful discussions with J. Sau. Part of this research was supported through ANR Grants No. ANR-11-JS04-003-01 and No. ANR-12-BS04-0016-03, an EU-FP7 Marie Curie IRG, and DOE under Contract No. DEFG02-08ER46482. Furthermore, we thank the Aspen Center for Physics for hospitality.

\setcounter{figure}{0}
\renewcommand{\thefigure}{S\arabic{figure}}
\setcounter{equation}{0}
\renewcommand{\theequation}{S\arabic{equation}}

\newpage
\begin{widetext}
\pagebreak

\section{Supplemental Material}

\subsection{The RSJ-model for the topological Josephson junction}

The voltage-biased RSJ-model~\cite{S-Shapiro-sup} is described by the equation
\begin{eqnarray}
V(t)\eq {\cal R}I_S(t)+\frac1{2e}\dot\varphi(t),
\end{eqnarray}
where $V(t)=V_{\rm dc}+V_{\rm ac}\cos\left(\Omega t\right)$ and $I_S(t)=I_J\sin\left[\varphi(t)/2\right]$. 

We rewrite
$$\varphi(t)=2k\Omega t +\frac{2eV_{\rm ac}}\Omega\sin\left(\Omega t\right)+\chi(t),$$
yielding
\begin{eqnarray}
\quad eV_{\rm dc}-k\Omega\eq e{\cal R}I_J\sum_mJ_{-m}\left(\frac{neV_{\rm ac}}\Omega\right)\sin\left[\left(k-m\right)\Omega t+\frac 12\chi(t)\right]+\frac1{2}\dot\chi(t).
\end{eqnarray}
Keeping only the slowly varying contribution $m=k$, we find
\begin{eqnarray}
\quad eV_{\rm dc}-k\Omega&\simeq&e{\cal R}I_JJ _{-k}\left(\frac{eV_{\rm ac}}\Omega\right)\sin\frac{\chi(t)}2+\frac1{2}\dot\chi(t)
\label{RSJ}
\end{eqnarray}
for $|eV_{\rm dc}-k\Omega|\ll eV_{\rm dc}$. For $k=0$, Eq. \eqref{RSJ} describes the supercurrent branch at low bias. The Shapiro steps at $eV_{\rm dc}=k\Omega$ are replicas of the supercurrent branch  \cite{S-quantronics} with a reduced maximal current $I_k=I_J|J _{k}(\alpha)|$, where $\alpha=eV_{\rm ac}/\Omega$.

From Eq. \eqref{RSJ}, we extract the characteristic time scale $\tau_R^{(k)}=1/[e{\cal R}I_J|J_k(\alpha)|]=1/(e{\cal R}I_k)$.
For $|(eV_{\rm dc}-k\Omega)\tau_R^{(k)}|\leq1$, we find the constant solution
$\sin\chi=(eV_{\rm dc}-k\Omega)\tau_R^{(k)}$, whereas, for $|(eV_{\rm dc}-k\Omega)\tau_R^{(k)}|>1$, integration of Eq.~\eqref{RSJ} over one period $T$ yields
\begin{eqnarray}
T= \frac{2\pi\tau_R^{(k)}}{\sqrt{\left[(eV_{\rm dc}-k\Omega)\tau_R^{(k)}\right]^2-1}}.
\end{eqnarray}
The dc current $I_{\rm dc}=\overline{I_S}$ is then given as
\begin{eqnarray}
I_{\rm dc}&\simeq& I_JJ _{-k}\left(\alpha\right)\overline{\sin \frac{\chi(t)}2}
=\frac1{e{\cal R}}\left\{eV_{\rm dc}-k\Omega-\frac12\overline{\dot\chi(t)}\right\},
\end{eqnarray}
yielding Eq. (10) in the main text. 

\subsection{Long-time properties of the topological Josephson junction in the limit $\tau_R\to\infty$}

In the presence of switching such that $\tau_s\ll\tau_R$, we can get analytic results for the situation of perfect voltage bias, ${\cal R}=0$. Using the long-time probabilities, cf.~Eqs (9) in the main text, we obtain
\begin{eqnarray}
\langle I(t)\rangle\eq\frac2\pi I_J\frac s{2-s}\sum_{n=-\infty}^\infty\frac1{1-4n^2}\sum_{k=-\infty}^\infty J_{-k}\left(2n\frac{eV_{ac}}\Omega\right)\cos\left[(2neV_{dc}-k\Omega)t+n{\varphi_0}\right],
\end{eqnarray}
where $\varphi_0$ is the initial phase.

In particular, the dc current reads
\begin{eqnarray}
\langle I_{\rm dc}\rangle\eq\frac2\pi I_J\frac s{2-s}\sum_{n=-\infty}^\infty\frac1{1-4n^2}\sum_{k=-\infty}^\infty J_{-k}\left(2n\frac{eV_{ac}}\Omega\right)\cos\left(n{\varphi_0}\right)\delta_{2neV_{dc},k\Omega}, 
\end{eqnarray}
consisting of a \lq\lq background\rq\rq\ current independent on the initial phase,
\begin{eqnarray}
\langle I_{\rm dc}^{(bg)}\rangle\eq=\frac2\pi I_J\frac s{2-s},
\end{eqnarray}
as well as delta functions at $2neV_{\rm dc}=k\Omega$ for $n\neq0$, with an amplitude dependent on the initial phase.

Similarly, we may also compute the finite-frequency noise,
\begin{eqnarray}
S(\omega)=\frac{i}{4\pi^2}I_J^2\frac {s^2}{(2-s)^2}\sum_{n,m,k,l;\pm}\frac{J_{-l}\left[(2m+1-ix)\alpha\right]J_{-(l+k)}\left[(2(m+n)+1-2ix)\alpha\right]}{(m-ix)(m+1-ix)(m+n-ix)(m+n+1-ix)}\frac{e^{in\varphi_0}\delta_{2neV_{dc},k\Omega}}{(2m+1-ix)eV-l\Omega\mp\omega}.
\end{eqnarray}
Here the \lq\lq background\rq\rq\ noise independent on the  initial phase reads
\begin{eqnarray}
S^{(bg)}(\omega)\eq\frac{iI_J^2{s^2}}{4\pi^2(2-s)^2}\sum_{m,l;\pm}\left[\frac{J_{-l}\left((2m+1-ix)\alpha\right)}{(m-ix)(m+1-ix)}\right]^2\frac{1}{(2m+1-ix)eV_{dc}-l\Omega\mp\omega}.
\end{eqnarray}
For $s\ll1$, the noise $S^{(bg)}(\omega)$ has peaks at $\omega=\pm(eV_{dc}-l\Omega)$. Namely,
\begin{eqnarray}
S^{(bg)}(\omega)\simeq\frac12I_J^2J_l^2(\alpha)\frac{seV_{\rm dc}/\pi}{[\omega\mp(eV_{dc}-l\Omega)]^2+(seV_{\rm dc}/\pi)^2},
\end{eqnarray}
for $|\omega\mp(eV_{dc}-l\Omega)|\ll eV_{\rm dc}$. As for the dc current, in addition, there are delta functions at $2neV_{\rm dc}=k\Omega$ for $n\neq0$, with an amplitude dependent on the initial phase.

The characteristic time scale for the current and noise to become independent of the initial phase is $\tau_R$. Thus, to obtain the shape of Shapiro steps as well as of the additional features in the noise spectrum, it would be necessary to determine the occupation probabilities $P(\varphi,t)$ and $Q(\varphi,t)$ in the presence of a finite series resistance ${\cal R}$. We leave this for further study.

\subsection{Comparison with the numerical results of Badiane {\it et al.}~\cite{S-Badiane2011}}

The current noise spectrum of a perfectly voltage-biased topological Josephson junction was considered earlier in Ref.~\cite{S-Badiane2011}. In that work, the problem of multiple Andreev reflections was solved numerically to obtain the frequency dependence of the current noise for a junction with a transmission probability $D$, at applied dc voltage $V_{\rm{dc}}$. In the limit of large transmission probability, the present work shows that the same current noise spectrum is described by Eq.~(14) of the main text. To compare the two approaches, we use Eq.~(14) with the switching probability $s$ as a free parameter to fit the numerical curves. The agreement is  very good. Some examples are shown in Fig. S1.

\begin{figure}[h]
\includegraphics[width=0.308\linewidth]{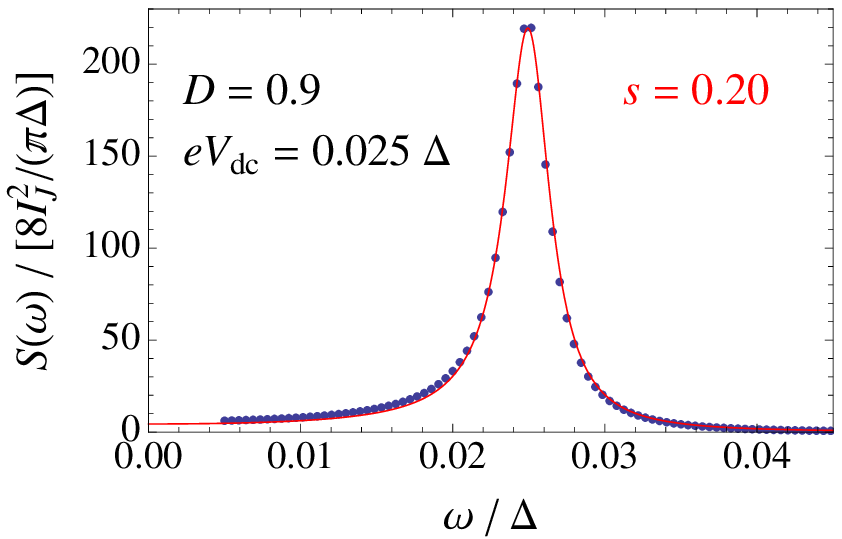}
\qquad\includegraphics[width=0.3\linewidth]{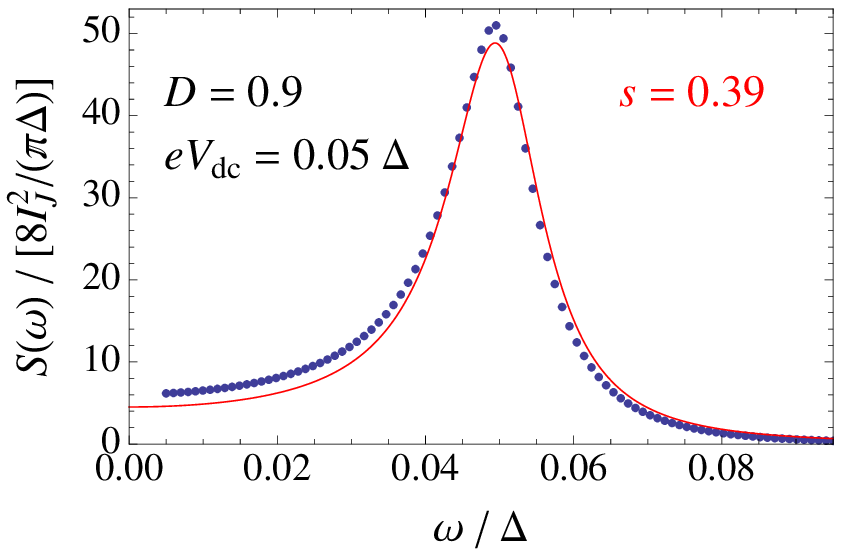}
\qquad\includegraphics[width=0.3\linewidth]{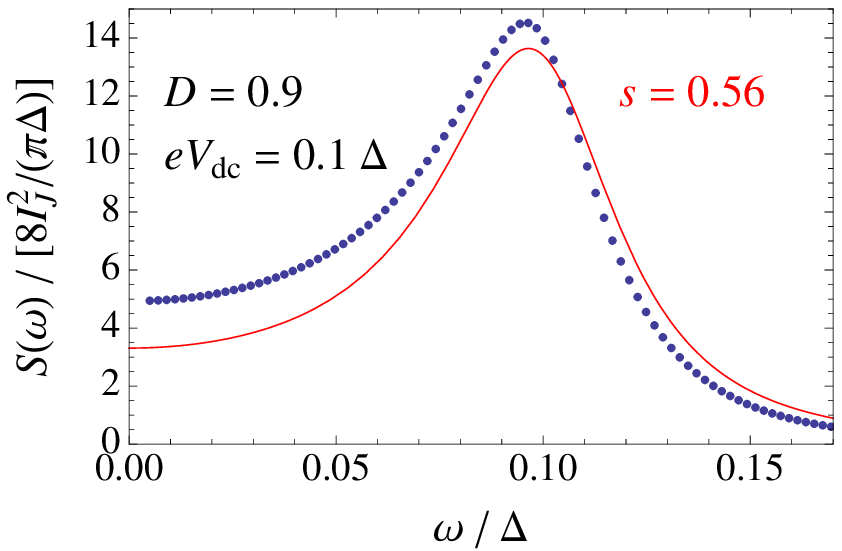}
\caption{
Current noise spectrum $S(\omega)$ as a function of the frequency $\omega$, for junctions with transmission probability $D=0.9$ and different applied dc voltages $V_{\rm{dc}}$. Dots: $S$ extracted by solving numerically the problem of multiple Andreev reflections~\cite{S-Badiane2011}. Lines: $S(\omega)$ obtained from fitting the numerical results with Eq.~(14) of the main text using $s$ as a free parameter.
}
\label{FigS2}
\end{figure}

The extracted switching probability may then be compared with the calculated switching probability as shown in Fig.~1 of the main text. In Fig.~S2, we present a more detailed comparison distinguishing different values of the transmission probability. As predicted in the present work, the curves all collapse when plotted as a function of the parameter $\lambda$, see main text. 
\begin{figure}[h]
\includegraphics[width=0.4\linewidth]{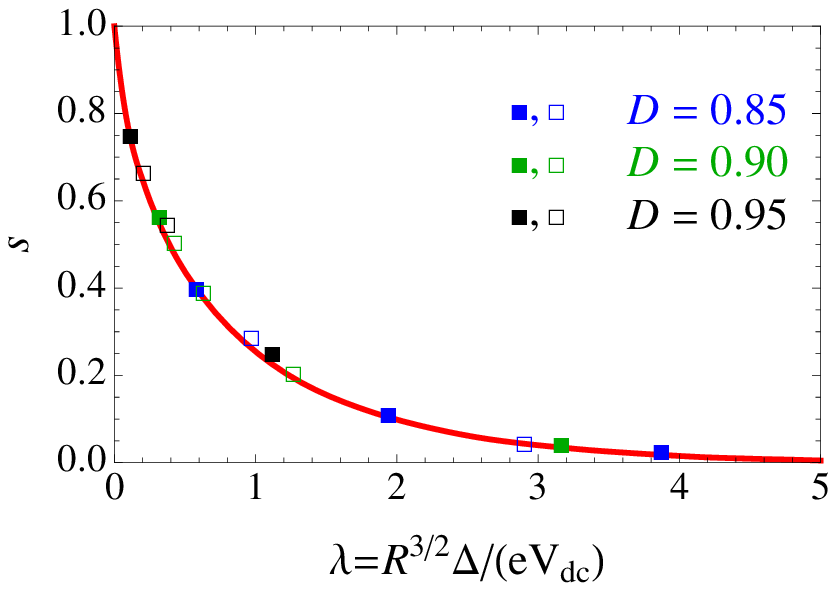}
\caption{
Switching probability $s$ as a function of the adiabaticity parameter $\lambda$. The line interpolates the values for $s$ found from a numerical solution of the Schr\"odinger equation with Hamiltonian \refe{eq:Hn} of the main text. Squares: $s$ extracted from the \lq\lq brute-force\rq\rq evaluation of the noise spectrum by solving numerically the problem of multiple Andreev reflections \cite{S-Badiane2011} and fitting the result by Eq.~(14) of the main text, for junctions with different transmission probabilities $D$. The filled squares correspond to the data shown in Fig.~1 of the main text.
}
\label{FigS2}
\end{figure}

\newpage
\end{widetext}


\begin{thebibliography}{99}
\bibitem{footnote2} 
The same physics also applies to  nontopological junctions if the gap between the bound states at $\varphi=(2n+1)\pi$  is much smaller than the gap to the continuum at $\varphi=2n\pi$ \cite{Billangeon,nonTopo}.
\end{thebibliography}
\end{document}